\documentclass{icrc}

\usepackage{times}
\usepackage{graphicx} 

\begin{document}
\title{Anisotropy analysis of EAS data in the knee region}
\author{S.V. Ter-Antonyan}
\affil{Yerevan Physics Institute, 2 Alikhanian
Br. Str., 375036 Yerevan, Armenia}

\correspondence{samvel@jerewan1.yerphi.am}

\firstpage{1}
\pubyear{2001}

\maketitle
\begin{abstract}
Based on MAKET-ANI EAS data the distributions of
equatorial coordinates of EAS core directions are obtained in the knee
region. Anisotropy of primary cosmic rays is displayed
only by declination equatorial coordinates ($\delta_a\simeq20^0\pm3^0$) at
primary energies more than 5-10 PeV. The fraction of anisotropic component 
turns out $\sim10\%$ in the knee region.
\end{abstract}

\section{Introduction}
The behavior of EAS size spectra in the knee region points out 
a possibility of a multi-component nature of the primary
nuclear flux \citep{JL, TG, PB, SBG, EW, TH}. 
As follows from \citep{SBG,TH,BONNY}, the best description of EAS size
spectra in the knee region may be achieved providing at least two
components in the primary cosmic ray flux. Confirmation of
the multi-component origin of cosmic rays can be obtain by   
investigations of distribution of EAS arrival directions.
Especially, it is interesting to measure of the anisotropy
in the vicinity of the knee region ($10^{15}-10^{16}$ eV) where
the
accuracies of EAS experiments in last years have significantly
increased. \\
In this paper, based on MAKET-ANI \citep{ANI} EAS data 
($\sim10^6$ events with $N_e>10^5$) the equatorial coordinate
distribution of EAS arrival directions are investigated. 
Violations of isotropy ($\sim5-10\%$) are obtained  
at energies $E>10$ PeV.

\section{EAS anisotropy}
Principal complications of the
investigation of the EAS ani- sotropy are:
\begin{enumerate}
\item EAS attenuation in the atmosphere (the intensity of shower
size $F(N_e,\theta)$ strongly depends on a zenith angle of incidence
($\theta$) at the given observation level);
\item dependence of the EAS size on the energy ($E$) and primary nucleus
($A$);
\item interruptions of exposition local time, which are inevitable in EAS
experiments.
\end{enumerate}
Factors 1-3 do not allow in principle to apply the traditional (on-off)
methods of measuring the cosmic-ray anisotropy and here, 
a simulation method is applied
taking into account above factors in the frameworks
of QGSJET interaction model \\
\citep{QGS} and predictions of the 2-component
origin of cosmic rays \citep{PB}. The test of this models by modern
EAS size spectra at different zenith angles and different
observation
levels one can find in papers \citep{TH,BONNY}.\\
Assume that the primary nuclear flux $I(E)$ consists of isotropic
$I_0(E)$
and anisotropic $\Im(E)$ components \citep{G} 
\begin{equation}
I(E)=I_0(E)+\Im(E,\alpha_a,\delta_a)\;\;,
\end{equation}
where $\alpha_a, \delta_a$ are equatorial coordinates of anisotropic
direction. 
\begin{figure*}[t]
\includegraphics[width=17.0cm,height=12cm]{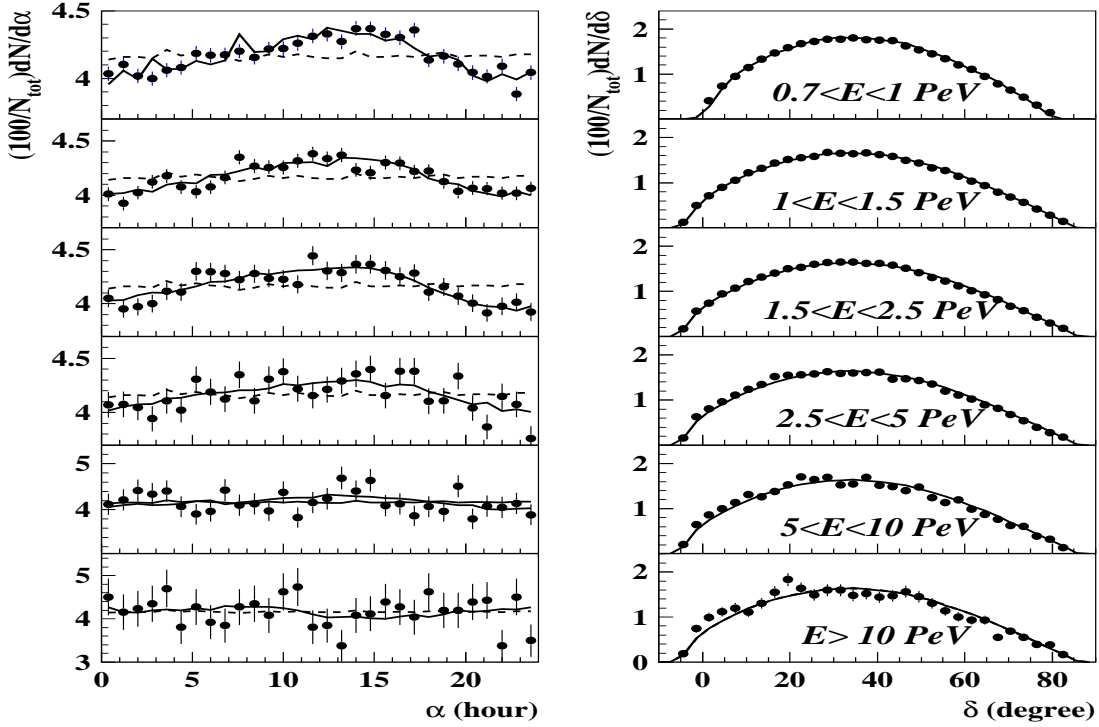} %
\caption{Equatorial coordinate distributions of all showers
at different primary energies ($E$).
Symbols are the Maket-ANI data. Solid lines are the
expected distributions obtained by simulations of the isotropic primary
component
(off-component) taking into account the local time of each detected
EAS event. Dashed lines are the isotropic components
at the uniform local time distribution of each EAS detected event.}
\end{figure*}
Let also the factor of anisotropy $\varepsilon=\Im(E)/I(E)$ is small 
($\varepsilon\ll 1$)
and does not distort the EAS zenith angular distribution due to
the Earth rotation and independence of EAS attenuation on the azimuthal
angle of incidence ($\varphi$).\\

Then, providing each detected EAS event from $I(E)$ flux with 
horizontal coordinates $\theta,\varphi$ and local time $\vec{T}$ 
by adequate simulated event but from isotropic 
$I_0(E)$ flux with corresponding coordinates $\theta^*,\varphi^*$ and
the same $\vec{T}$ one can compare (after
transformation $\theta,\varphi,\vec{T}$ and $\theta^*,\varphi^*,\vec{T}$
$\Rightarrow$ 
$\alpha,\delta$ and $\alpha^*,\delta^*$) the obtained equatorial
coordinate
distributions of real EAS data and isotropic simulated data.
Evidently, if the anisotropic
part of intensity $\Im(E,\alpha_a,\delta_a)=0$ then the distribution 
obtained from the experiment must overlap with the simulated one. Any
considerable
discrepancies (out of statistical errors) will point out to the 
existence of anisotropy.\\
The presented method completely solves the problems 1 and 3,  however, the
same normalization of real and simulated distributions
(due to a flux $\Im(E,\alpha_a,\delta_a$) is unknown) does not give a 
direct possibility to determine the absolute intensity of
the anisotropic component.\\

Taking the above into account,
each detected EAS event is characterized by a 9-dimensional vector 
$\vec{D}(E,\theta,\varphi,\vec{T}(y,m,$ $d,h,u,s))$, 
where $E\equiv E(N_e,\theta)$ is the evaluation of a primary energy
on the basis of the measured EAS size $N_e$ and zenith angle $\theta$, 
$\vec{T}(y,m,d,h,u,e)$
is a local time vector with components $y$~-year, $m$~-month,
$d$~-day, 
$h$~-hour, $u$~-minute, $e$~-second of a detected event. Here we used
the vector definition only for briefness and convenience. \\
\begin{figure*}[t]
\includegraphics[width=17.0cm,height=12cm]{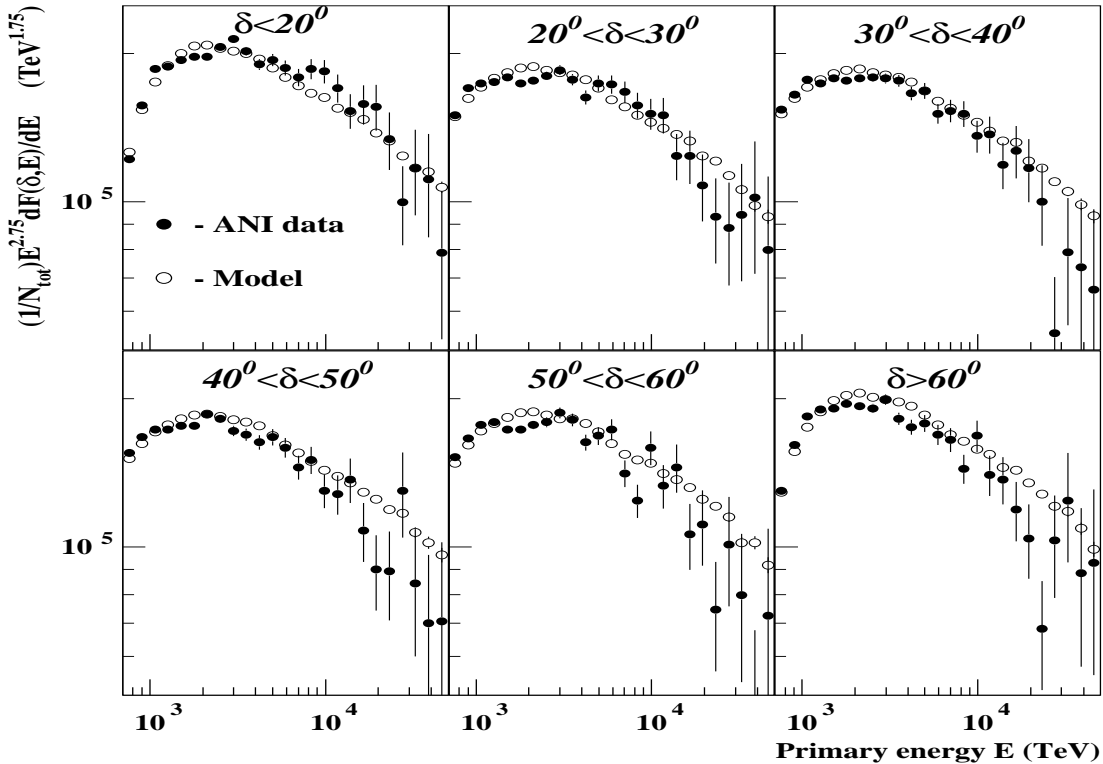} %
\caption{Detected normalized all-particle primary energy spectra at
different equatorial
declination $\delta$-coordinates. Black symbols
correspond to ANI EAS data. Open 
circles are the expected energy spectra of an isotropic component 
(off-component) according to \citep{PB} model.}
\end{figure*}

The estimation of the primary energy $E$ by the measured EAS size $N_e$
is performed by inverse interpolation of function 
$\overline{N_{e}}\equiv \overline{N_{e}}(E,\overline{A},\theta)$
at average primary nucleus $\overline{A}(E)=$  $\exp(\overline{\ln 
A})$. $\overline{A}$ values are determined by  
2-component primary spectra \citep{PB}. A tabulated function 
$\overline{N_{e}}(E,A,\theta)$ at given $E$, $A$, $\theta$
parameters
of a primary nucleus was preliminary calculated by means of 
EAS simulation using the  
CORSIKA code \citep{COR} at QGSJET interaction model
\citep{QGS}. Some details of this simulation one can find in the 
paper \citep{TH}. Statistical errors did not exceed $3-5\%$
and the high altitude location ($700$g/cm$^2$) of the ANI experiment
\citep{ANI}
allowed to obtain $\Delta E<20\%$.

Because the exposition time of the MAKET-ANI array is not
continuous
and the EAS zenith angular distribution does not adequate to the angular
distribution of primary nuclei
(factors 1,3 above), we created
$i=1,\dots,m$ simulated 9-dimensional vectors 
$\vec{R}_i(E^*_i,\theta^*_i,\varphi^*_i,\vec{T})$ 
for each detected vector $\vec{D}$ which differ from
$\vec{D}$ only by simulated primary energy ($E^*$), zenith
($\theta^*$) and azimuth ($\varphi^*$) angles.\\
The values of angles $\theta^*$ and $\varphi^*$ are simulated according to
distributions: $dF/d\cos\theta=\cos\theta$ and $d\Phi/d\varphi=1/2\pi$
respectively. The energy ($E^*$) of simulated events is obtained from 2-component
primary energy spectra \citep{PB,BONNY} at additional condition
of detected events $N_{e}^*(E^*,A^*,\theta^*)>10^5$.

Transformation of $\vec{D}$ and $\vec{R}_i$ vectors
to simple 3-dimensional vectors $\vec{d}(E,\alpha,\delta)$ and
$\vec{r}_i(E^*,\alpha^*,\delta^*)$ with 
$\alpha$ -right ascension and $\delta$ -declination
equatorial coordinates are performed by astronomical soft 
\citep{SK}. 

\section{Results}
The results of investigation of the EAS anisotropy in the knee region
by ANI EAS data are shown in Fig.~1,2. Equatorial coordinate distributions
$dN/d\alpha$ and $dN/d\delta$ of all showers 
($N_{tot}=10^6$ events at $N_e>10^5$) at six primary energy 
($E$) intervals
are given in Fig.~1. Symbols are the Maket-ANI data
$dN/d\alpha$ and $dN/d\delta$ in terms of ($100/N_{tot}$). Solid
lines are the expected adequate distributions 
$dN/d\alpha^*$ and $dN/d\delta^*$ obtained by simulations of
isotropic primary component (off-component). Local times 
$\vec{T}$ of simulated events
and detected EAS events are equal. The total number of events
is the same ($N_{tot}=N^{*}_{tot}$). 
Dashed lines reflect the expected behavior of the isotropic
component at an uniform local time distribution of each EAS detected
event.\\
It is seen, that the distribution of a $\delta$-coordinate at all 
primary energies is practically  
independent on a local time distribution. Moreover, a good
agreement of expected and detected distributions for equatorial
$\alpha$-coordinate is observed in the energy range $E>0.7$ PeV.\\
However, the discrepancies of detected and expected distributions 
of the equatorial $\delta$-coordinate increase with the energy. 
Strong disagreement is observed at $E>10$ PeV and
$\delta\simeq20^0$. It is necessary to note, that the real disagreement
is a little larger than it is seen in Fig.~1 because EAS and
simulated data have the same normalizations.
Further investigations of the $\delta$-coordinate distribution
are given in Fig.~2 where detected normalized all-particle
primary energy spectra at different declinations ($\delta$) are presented. 
Black symbols are ANI data.  The open circles are expected
energy spectra of isotropical simulated components (off -
component) with 50 times higher statistics
($N^{*}_{tot}=5\cdot10^7$).\\
Spectral shape in $E<1-2$ PeV range for all declination
intervals is determined by the lower limit of the detected EAS size 
($N_e>10^5$). However, the observed fine structure of 
all-particle spectra at $E>2$ PeV and declinations $\delta<30^0$ and
$50<\delta<60$ can hardly  be explained by trivial statistical
fluctuations.
All data in Fig.~1,2 are obtained in the framework of QGSJET
interaction model \citep{QGS} and the 2-component origin of cosmic rays 
\citep{PB,BONNY}.
\balance 

\section{Conclusion}
From above analysis it follows that there is a partial anisotropic flux
with a
declination coordinate equal to $\delta_a=20^0\pm3^0$ at primary energies
$E>10$ PeV.
The distribution of equatorial $\alpha$-coordinates agrees with the
hypothesis of the isotropic distributions of cosmic ray.\\
It is interesting to note, that the nearest neutron stars PSR0950 and
PSR1133 disposed at $l\simeq30-100\;ps$ distance from the Earth have
declination coordinates $\delta=30.6^0$ and $\delta=27.45^0$ respectively.
These values are close to the results obtained here ($\delta_a\sim20^0$). 
Are these neutron stars a reason of anisotropic flux or not
can be elucidated only with significantly large EAS statistics.

\begin{acknowledgements}
It is a pleasure to thank Ashot Chilingaryan for many
helpful discussions about these matters. I thank  
Lilith Haroyan and Hamlet Martirosyan for their assistance during the
work. I thank the members of the MAKET-ANI group for providing EAS
data.\\
The work has been partially supported by the research grant N 00-784 
of the Armenian government,
NATO NIG-975436 and CLG-975959 grants and ISTC A216 grant.\\
This publication is based on experimental results of the ANI
collaboration.
MAKET ANI installation has been set up as a collaborative project of
the Yerevan Physics Institute, Armenia and the  Lebedev Institute, 
Russia. Continuous contribution of Russian colleagues 
is thankfully acknowledged.
\end{acknowledgements}

\end{document}